\documentclass[onecolumn,authoryear]{els-mrw} 

\usepackage{amsmath,amssymb,amsfonts,amsthm,makeidx,graphicx}
\usepackage{txfonts}
\usepackage{helvet}
\usepackage{hyperref}
\usepackage{sidecap}

\newcommand{\nplanets}[1]{5700#1}

\begin{document}

\chapter{Exoplanet Demographics: Physical and Orbital Properties}\label{chap1}

\author[1]{Ryan Cloutier}%

\address[1]{\orgname{McMaster University}, \orgdiv{Department of Physics \& Astronomy}, \orgaddress{1280 Main St W, Hamilton, ON L8S 4L8, Canada}}


\maketitle

\begin{abstract}[Abstract]
The discovery of over \nplanets{} exoplanets has led to a boom in the field of exoplanet demographics over the past decade. Led by swaths of exoplanet discoveries from NASA's Kepler space mission, astronomers have been conducting statistical studies of the exoplanet population in search of trends in various planetary and host stellar parameters. These investigations are informing our understanding of how planets form and evolve, thus putting our solar system into a galactic context. In this chapter, we review many of the major features uncovered in the distributions of physical and orbital parameters of known exoplanets including the Radius Valley, the Neptunian Desert, the Peas in a Pod pattern, dynamical properties that point toward likely formation/migration mechanisms, as well as trends with host stellar parameters such as the time-evolution of exoplanetary systems and the search for planets within the Habitable Zone. The overarching theme is that exoplanetary systems exhibit an incredible diversity of planet properties and system architectures that do not exist within our own solar system. A promising future awaits the field of exoplanet demographics with increasingly deep investigations planned following the launch of numerous dedicated space telescopes over the coming years and decades.
\end{abstract}

\begin{glossary}[Glossary]
\term{Astronomical unit} the average Earth-Sun distance throughout the Earth's orbit. Commonly abbreviated au, it is equal to $1.5 \times 10^8$ km. \\
\term{Apogee} the farthest distance of an orbiting body from its host.\\
\term{Exoplanet} a planet that orbits a star other than the Sun.\\
\term{Equation of state} a function that expresses the relationship between a material's thermodynamical state variables temperature, pressure, and density. \\
\term{Greenhouse effect} heating of a planetary surface by the absorption and re-radiation of outgoing thermal radiation from the planet's surface by greenhouse gases such as water, methane, and carbon dioxide. \\
\term{Gyrochronology} a technique used to infer the age of an individual star from the measurement of its rotation period. \\
\term{Habitable zone} the range of circumstellar orbital distances wherein a terrestrial planet is expected to be able to sustain liquid water on its surface. Many definitions of the circumstellar habitable zone exist in the scientific literature, all of which are conditioned on a set of model assumptions, in particular, regarding the bulk properties and atmospheric composition of the planet. \\
\term{Hill radius} the radial distance around a planet inside of which the planet's own gravitational potential dominates the orbit of a test particle (compared to the gravitational potential of the star). \\
\term{Kepler dichotomy} the division of exoplanetary systems with single transiting planets versus many planets that suggests the existence of two planet populations with different intrinsic multiplicities or mutual inclinations. \\
\term{Kepler space telescope} NASA's first transiting exoplanet survey mission that was designed to search for Earth-sized exoplanets in the habitable zones of other stars. Kepler discovered more than 2700 confirmed exoplanets and consequently revolutionized the field of exoplanet demographics. \\
\term {Kraft break} the effective temperature that separates stars that do and do not harbour deep convective envelopes. Convection is needed to drive large scale magnetic dynamos, which in term affect stellar rotation such that stars hotter than the Kraft break at 6200 K exhibit rapid rotation velocities. \\
\term{Mass-radius relation} the dependence of planetary mass on planetary radius. This could refer to the empirical distribution of planetary masses and radii or to theoretical calculations of planetary masses and radii for a particular bulk composition and irradiation temperature. \\
\term{Multis} short-form that refers to planetary systems with more than one known planet. \\
\term{NASA Exoplanet Archive} an online resource of collated exoplanetary and stellar parameters, which also serves as an access point to public datasets and analysis tools that have enabled exoplanet demographics studies by the exoplanet community at large. \\
\term{Neptunian desert} a wedge-shaped region of the exoplanetary radius-period and mass-period spaces marked by dearth of Neptune-sized planets with short orbital periods and believed to be the result of an atmospheric mass loss process. \\
\term{Peas in a Pod pattern} systems of tightly-packed inner planets with $\geq 4$ planets have architectures that exhibits correlated sizes, masses, and regular orbital spacings that resemble literal peas in a pod. \\
\term{Perigee} the closest distance of an orbiting body from its host.\\
\term{Radius Valley} the bimodal distribution of close-in small planet radii with a dearth of planets between $\sim 1.6-1.9\ R_\oplus$. Is thought to mark the distinction between terrestrial versus volatile-rich planetary compositions. \\
\term{Red dwarf stars} stars whose internal energy transport mechanisms are either Sun-like (see below) or fully convective {\it and} have masses that are $\lesssim 0.6\, M_\odot$, making their surfaces cooler and therefore their colours redder than the Sun. \\
\term{Refractories} chemical species characterized by their tendency to resist phase changes at all but extremely hot temperatures ($\gtrsim 1000$ K). Common examples include silicates, iron, and carbonaceous chondrites. \\
\term{Singles} short-form that refers to planetary systems with one known planet. \\
\term{Sub-Neptunes} a catch-all term referring to exoplanets with sizes smaller than Neptune and with bulk compositions that are volatile-rich and hence, are lower density than super-Earths of the same mass. Sub-Neptune may be further categorized depending on the phase and mixing ratios of the volatiles of interest (e.g. H-He-enveloped, water world, etc.). \\ 
\term{Sun-like stars} stars whose internal energy transport mechanisms are radiative in their nuclear-fusing cores and convective in their outer envelopes. These stars typically have masses in the range $0.6-1.3\, M_\odot$. \\
\term{Super-Earths} a class of exoplanets characterized as having a terrestrial composition while being larger/more massive than the Earth. \\
\term{Systems of Tightly-packed Inner Planets (STIPs)} common architectures of multi-planet systems characterized by their compact orbital spacings at small orbital distances from the central host star. \\
\term{Volatiles} chemical species characterized by their low sublimation temperatures. Common examples include hydrogen, helium, water, methane, ammonia, and carbon dioxide. \\
\end{glossary}

\begin{glossary}[Nomenclature]
\begin{tabular}{@{}lp{34pc}@{}}
au & Astronomical unit\\
C/O & Carbon-to-Oxygen ratio \\
CMF & Iron Core Mass Fraction \\
EOS & Equation of State \\
ESA & European Space Agency \\
GH & Greenhouse \\
Gyrs & Giga-years (i.e. a billion years) \\
H-He & Hydrogen and Helium (refers to a mixture of these two gases) \\
HEM & High-Eccentricity Migration \\
HJ & Hot Jupiter \\
HZ & Habitable Zone \\
K/O & Potassium-to-Oxygen ratio \\
K2 & Kepler's ``second light'' mission (pronounced kay-two)\\
Kepler & the Kepler Space Telescope\\
$\odot$ & the subscript denoting the Sun (i.e. $M_\odot=$ the solar mass)\\
$\oplus$ & the subscript denoting the Earth (i.e. $M_\oplus=$ the Earth's mass)\\
Myrs & Mega-years (i.e. a million years) \\
NASA & National Aeronautics and Space Administration\\
PiaP & Peas in a Pod pattern \\
RV & Radial Velocity\\
STIP & Systems with Tightly-packed Inner Planets \\
TESS & Transiting Exoplanet Survey Satellite\\
USP & Ultra-Short Period planet \\
WMF & Water Mass Fraction \\
XUV & X-ray and Ultraviolet \\
\end{tabular}
\end{glossary}

\begin{BoxTypeA}[chap1:box1]{Key points}
\begin{itemize}
\item {\bf Exoplanet demographics} is the study of the exoplanet population. Trends with planetary and/or stellar parameters inform our understanding of how planets form and evolve.
\item The exoplanet population exhibits an {\bf incredible diversity} that features many types of planets that do not exist in our own solar system.
\item The {\bf Radius Valley} is a dearth of planets at $\sim 1.8\, R_\oplus$ in the exoplanet size distribution that separates terrestrial super-Earths from the larger and lower density sub-Neptunes.
\item The empirical {\bf mass-radius relation} of exoplanets reveals different regimes of terrestrials, volatile-rich worlds, and gas giants, each with their own characteristic dependence of size and density on planetary mass.
\item The {\bf Neptunian Desert} is a feature in the exoplanet size-mass-period distribution characterized by the lack of Neptune-sized objects at close orbital periods, believed to by sculpted by processes of XUV-driven atmospheric escape and tidal distribution.
\item Systems of tightly-packed inner planets with more than four planets exhibit {\bf correlated sizes, masses, and orbital spacings} that resemble literal peas in a pod.
\item Given the prevalence of STIPs discovered by Kepler, there exists an apparent excess of single transiting systems that suggests the existence of {\bf two planet populations} with different multiplicities or mutual inclinations.
\item The orbital spacings of STIPs tend to cluster around a value that is just beyond the criterion for dynamical stability, which suggests that {\bf significant dynamical processing likely acts on young planetary systems} and leaves beyond system architectures that are at the edge of stability.
\item The exoplanet population exhibits a much {\bf broader distribution of orbital eccentricities} compared to the solar system's planets, with multi-planet systems showing systematically lower eccentricity values than single planet systems, which is consistent with dynamical stability considerations.
\item The exoplanet population exhibits a much {\bf broader distribution of orbital obliquities} compared to the solar system's planets, including numerous examples of hot giant planets with nearly polar orbits as well as retrograde orbits.
\item {\bf Planets' sizes and orbits are expected to evolve over time with particularly rapid evolution taking place during the earliest stages of the system's lifetime. However, most stars are highly active at young ages, which stymies exoplanet detections and makes comparisons between young to old planetary systems difficult.}
\item The {\bf circumstellar habitable zone} refers to the range of orbital distances within which an Earth-like planet with an Earth-like atmosphere is expected to be able to sustain liquid surface water and therefore be considered potentially habitable to life like our own.  
\end{itemize}
\end{BoxTypeA}

\section{Introduction}\label{chap1:sec1}
Extra-solar planets--or exoplanets--are planets that orbit stars other than our Sun. Since the discovery of the first exoplanet orbiting a Sun-like star in 1995 \citep{Mayor_1995}, astronomers have uncovered a wild diversity of exoplanets. Many of the more than \nplanets{} confirmed exoplanets are unlike anything that exists in our own solar system with distinct masses, radii, orbital parameters, and host star properties. Astronomers use this information to search for trends in the distributions of exoplanetary and host stellar parameters to inform our understanding of the planet formation and evolutionary processes, and their prevalences throughout our galaxy.

\begin{figure}
\centering
\includegraphics[width=1\hsize]{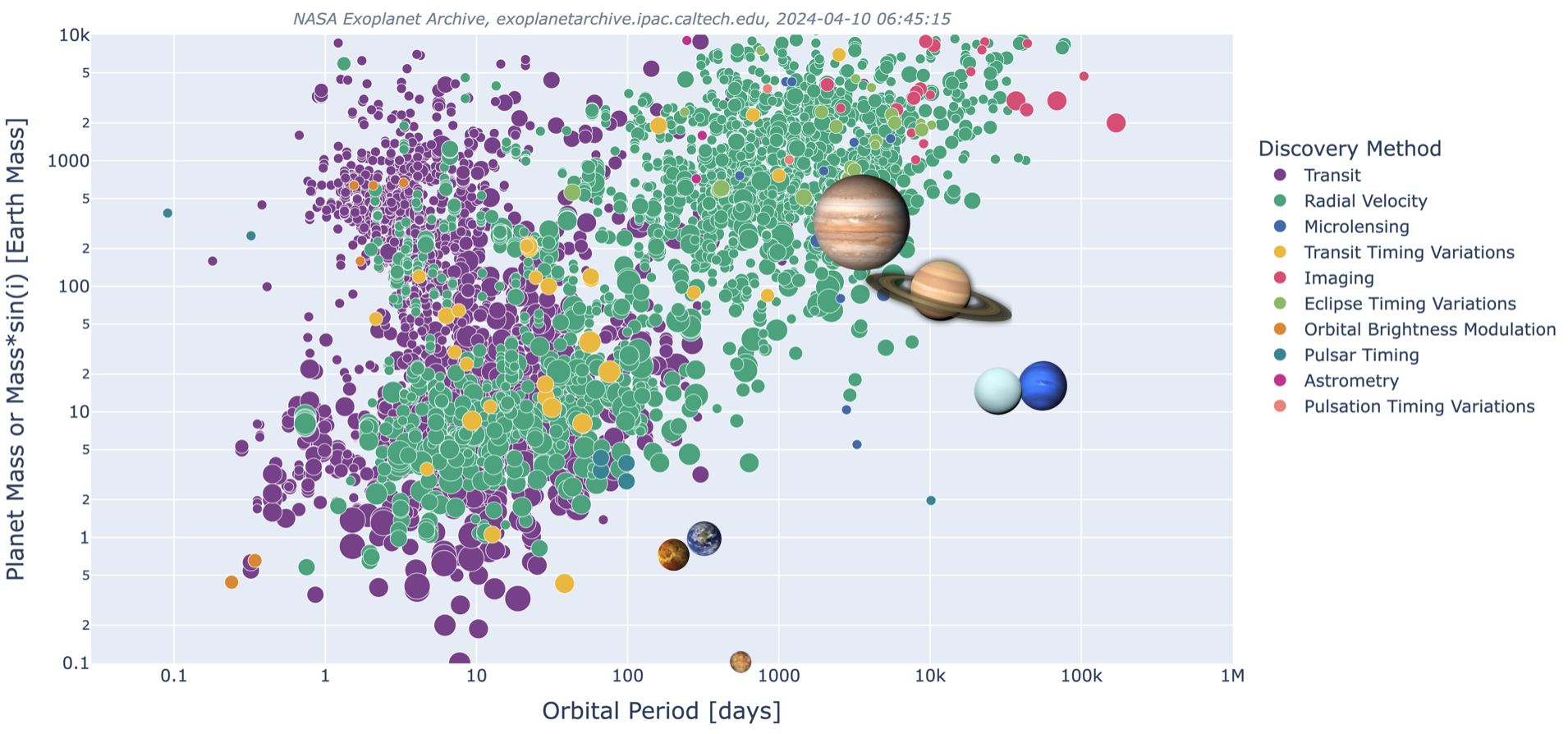}
\caption{Orbital periods and planetary masses, or minimum masses where available, of confirmed exoplanets compared to the planets in the solar system (modulo Mercury). The exoplanet marker sizes represent the number of known planets in the system with larger markers representing higher planet multiplicity. The sizes of the solar system planets do not carry any particular meaning. The marker colours depict the planet's discovery method. Data from the NASA Exoplanet Archive, whose interactive plotting tool was used to construct the plot.}
\label{fig:demo}
\end{figure}

The population of confirmed exoplanets is depicted in Figure~\ref{fig:demo} revealing planets' orbital periods, planetary masses\footnote{Planetary minimum masses are reported in place of planet masses if the mass measurement was obtained using the radial velocity discovery method.}, 
discovery methods, and the multiplicities of known planets in each planetary system. This quintessential figure underpins the field of {\it exoplanet demographics}: the statistical study of population-level features in the distribution of exoplanets' physical and orbital parameters, as well as correlations with the properties of their host stars. Five of the most prominent exoplanet discovery methods are discussed in detail in subsequent chapters of this volume of the {\it Encyclopedia of Astrophysics} (e.g. Radial Velocity, Transit/Transit Timing Variations, Direct Imaging, Astrometry, and Gravitational Microlensing). Exoplanet discoveries using these methods are shown in Figure~\ref{fig:demo} along with a smattering of discoveries from less prominent methods. An impressive variety of exoplanet subpopulations are evident in Figure~\ref{fig:demo}, many of which are easily distinguishable by their discovery method due to the inherent biases of individual methods that favour particular stellar and planetary properties.
Many planet types have been assigned a unique nomenclature that tend toward qualitative descriptions of their properties rather than to strict definitions based on explicit parameter ranges. Major subpopulations that emerge in Figure~\ref{fig:demo} include \\

\begin{itemize} 
\item \textbf{Hot Jupiters (HJ)}: gas giant planets with orbital periods $\lesssim 10$ days and hence high irradiation temperatures due to their close proximity to their host stars. The majority of known HJs exist as the sole known member of their respective planetary systems, suggesting that HJ formation and migration is not reliant on gravitational interactions with other planets.  
\item \textbf{Warm/cold gas and ice giants}: gaseous planets with masses $\gtrsim 10\, M_\oplus$ and orbital periods between $\sim 10-100$ days and $\gtrsim 100$ days, respectively. In many ways, the physical and orbital properties of these planets make them analogous to the gas and ice giants in our solar system. 
\item \textbf{Super-Earths and sub-Neptunes}: planets with masses and sizes between those of the Earth and Neptune\footnote{Neptune's mass and radius are $17\, M_\oplus$ and  $3.9\, R_\oplus$, respectively.} over a range of orbital periods from $\sim 1-1000$ days. These planets dominate the set of confirmed exoplanets and are commonly found in multi-planet systems.
\item \textbf{Ultra-short period planets (USP)}: planets with orbital periods of less than one day. USPs can be further subdivided into terrestrial and gas giant USPs, which are commonly referred to as lava worlds and ultra-hot Jupiters, respectively.
\item \textbf{Exo-Earths and sub-Earths}: likely terrestrial planets with sizes $\lesssim 1.25\ R_\oplus$. These planets are commonly found in multi-planet systems, albeit at short orbital periods ($\lesssim 10$ days), which make them distinct from the similarly-sized terrestrial planets in the inner solar system. \\
\end{itemize}

A striking feature of the known exoplanet population in mass-period space is how little overlap there is with the planets in our solar system. While we do have examples of exo-Jupiters and exo-Saturns, analogs to the solar system's ice giants and inner terrestrial planets have yet to be discovered. We note however that this does not imply that these planets do not exist around other stars. Rather, this discrepancy likely arises from the limited sensitivity of our current discovery technologies to these types of planets.
It is important to note that the apparent dissimilarities between the solar system's planets and exoplanets may be softened in certain parameter spaces. For example, while confirmed exo-Earths and sub-Earths have orbital periods that are commonly much less than 365 days, if one compares the levels of irradiation that known exo-Earths receive from their host stars to the irradiation that the Earth receives from the Sun, there are numerous examples of exo-Earths that receive Earth-like irradiation because they orbit stars that are much less luminous that the Sun. However, it remains true that the solar system's architecture appears very different from the known exoplanet population.

Not only do exoplanetary parameters differ from those in our solar system, exoplanets also orbit a wide variety of stellar types that influence their formation conditions and evolutionary pathways. We have detected planets around massive stars $>2\, M_\odot$, whose high luminosities drive extreme heating on close-in planets to temperatures that exceed the surface temperature of some stars $\gtrsim 2800$ K. There is also a plethora of confirmed planets orbiting low mass stars whose stellar masses extend down to $0.08\, M_\odot$; the lowest mass objects that are capable of undergoing nuclear fusion in their cores and therefore be considered bona-fide stars. This diversity reveals that the planet formation process is largely ubiquitous throughout the galaxy, while the bulk properties and relative frequencies of different planet types around different stellar types lends crucial insights into how planetary formation and evolutionary processes are influenced by the host star. Looking deeper, even more exotic types of planet hosts have been discovered including evolved giant stars and ultra-dense stellar remnants such as white dwarfs and pulsars; the degenerate cores of stars that are leftover after their progenitor stars have lived out their lifetimes. Approximately twenty circumbinary planets that orbit a binary star system instead of a single star have also been discovered. Even free-floating planetary-mass objects (a.k.a. rogue planets) that are not gravitationally bound to a star or star system have been discovered and may represent the outcome of dynamical interactions that eject planets from their native planetary systems.

In this chapter on {\it Exoplanet Demographics: Physical and Orbital Properties}, we will explore some of the major features that have emerged in the exoplanet population, as well as some of the leading theoretical ideas that have been proposed to explain the origins of these phenomena.

\subparagraph{NASA Exoplanet Archive} \label{chap1:sec1:subsec1}
Keeping up with the whirlwind of planet discoveries and planet parameter refinement is no easy task given the pace that the field moves at. Fortunately, the National Aeronautics and Space Administration (NASA) Exoplanet Archive\footnote{Operated by the California Institute of Technology with support from NASA through the Exoplanet Exploration Program.} is an online database of exoplanetary parameters, host stellar parameters, and exoplanet discovery/characterization data. The NASA Exoplanet Archive serves as a resource of collated parameters and datasets that are retrieved from literature sources on an ongoing basis and is widely used by the community for demographics studies. Indeed, the overview of the known exoplanet population presented in Section~\ref{chap1:sec1} was based on data retrieved from the NASA Exoplanet Archive.

The NASA Exoplanet Archive is an indispensable tool that includes extensive data tables of confirmed and candidate planets from a variety of surveys, plus interactive plotting, analysis, and model fitting tools. These resources are all publicly available at this \href{https://exoplanetarchive.ipac.caltech.edu/}{url}.

\subparagraph{Kepler space telescope} \label{chap1:sec1:subsec2}
Although many ground and space-based facilities have contributed to our understanding of exoplanet demographics, no singular facility has been as impactful on the field as NASA's Kepler space telescope. This warrants some forewords on this revolutionary mission, which is frequently referenced throughout this chapter.

The Kepler space telescope--or Kepler for short--is a now decommissioned NASA space mission whose primary science goal was to detect Earth-sized planets in the habitable zones of (primarily) Sun-like stars. Kepler was designed to find exoplanets using the transit method by searching for periodic dimming events that occur when a planet transits in front of its host star, blocking a small fraction of the star's light that may be detectable by monitoring the star's brightness over time.
Kepler was launched on March 7, 2009 into an Earth-trailing heliocentric orbit and operated for a total of nine years in one of two distinct modes: the primary Kepler mission and the ``second light" K2 mission. During its primary mission (2009-2013), Kepler stared at approximately 150,000 stars over a 115 deg$^2$ patch of sky in the Cygnus-Lyra region located above the plane of the Milky Way\footnote{Kepler's 115 deg$^2$ field-of-view is equivalent to about 0.27\% of the entire sky.}. As of April 2024, 2742 confirmed exoplanets have been discovered in Kepler's primary mission, plus an additional 1983 planet candidates. This multitude of exoplanet discoveries transformed the field of exoplanet demographics by enabling statistical studies of planets' physical and orbital parameters. 

Following the failure of two of the Kepler spacecraft's reaction wheels between 2012-2013, which prevented the telescope from maintaining the precise pointing needed for planet hunting, the Kepler mission's observing strategy was reimagined into a ``second light" mission that came to be known as K2. By leveraging radiation pressure from the Sun, NASA engineers successfully repurposed the Kepler space telescope to observe fields in the ecliptic plane for approximately 83 days at a time. This reimagining of the Kepler mission into K2 began operating in 2014. Data from the K2 mission suffered elevated systematic noise compared to the primary mission due to its comparatively imprecise pointing precision that required periodic thruster firing every 12 hours to correct for the telescope's drift. While this pointing noise was ultimately detrimental to K2's ability to detect new small exoplanets, K2's focus on multiple fields that spanned the ecliptic plane greatly expanded upon Kepler's target list and, in particular, provided a large sample of small red dwarf stars around which habitable zone planets could be detected. As of April 2024, K2 has discovered 575 confirmed exoplanets, plus an additional 977 planet candidates.

The K2 mission, and therefore the Kepler mission as a whole, was deactivated on November 15, 2018. The Kepler mission leaves behind a substantive legacy of planet discoveries, many of which are still being validated and debated today, as well as continued demographics studies that have produced many of the major of features in the exoplanet population that we will discuss in the forthcoming sections.

\section{Population-level trends in planets' physical parameters}\label{chap1:sec2}
\subsection{Radius Valley}\label{chap1:sec2:subsec1}
\begin{figure}
\centering
\includegraphics[width=0.6\hsize]{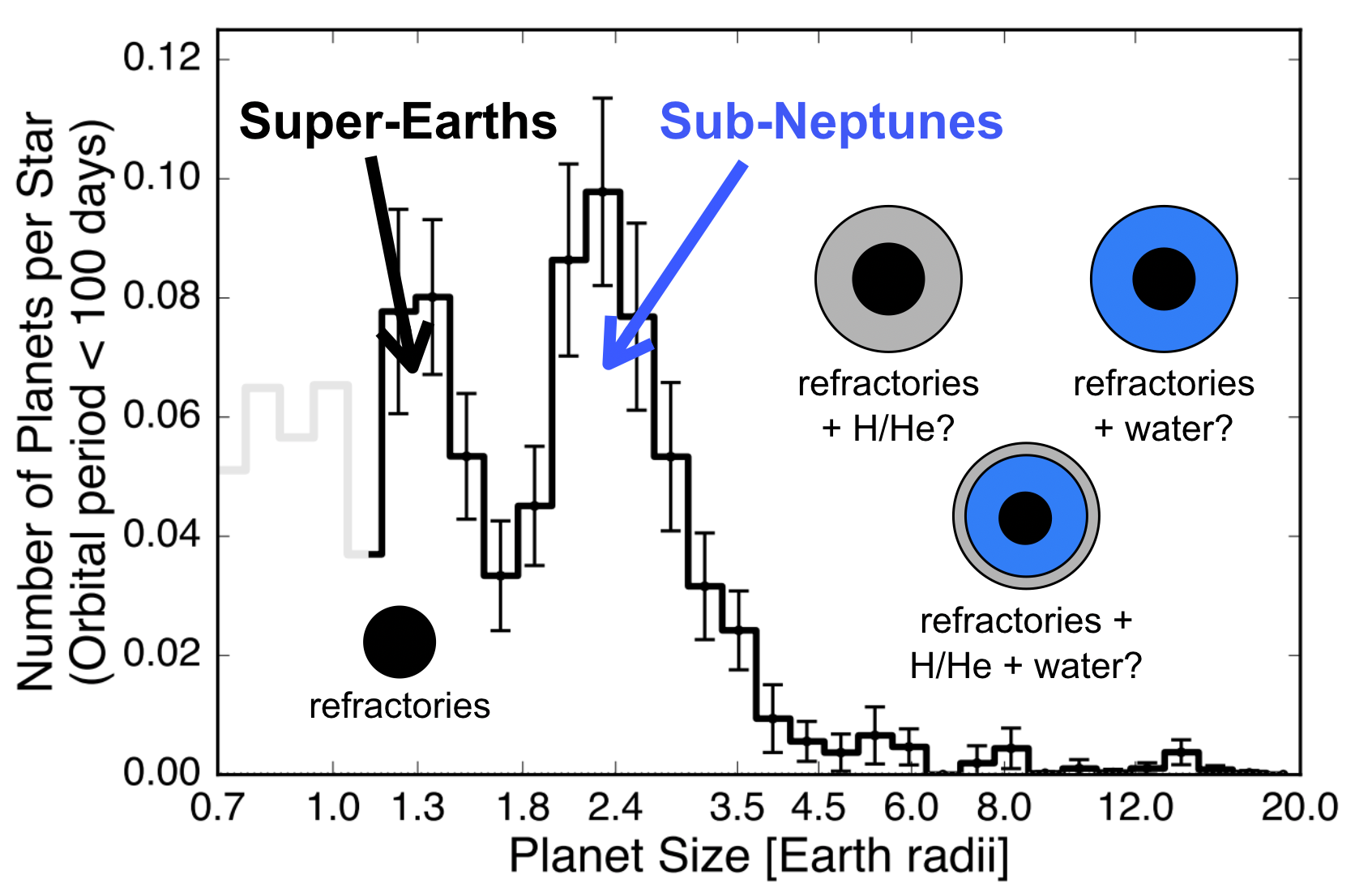}
\caption{The distribution of close-in planet radii exhibits a bimodality for sub-Neptune-sized planets ($\lesssim 4\, R_\oplus$) known as the Radius Valley. The Radius Valley distinguishes  small super-Earths, whose bulk compositions are composed of refractory materials such as iron and silicates, from the larger sub-Neptunes that require a substantial volatile mass fraction to explain their observed radii and masses. These volatiles likely include H-He gas and various phases of water, but the exact nature and diversity of sub-Neptune compositions remains a major open question in the field. Modified from \cite{Fulton_2017}.}
\label{fig:radval}
\end{figure}

One of the most important discoveries in the field of exoplanet demographics over the past decade has been the discovery that the distribution of close-in planet radii exhibits a bimodal distribution separated by a dearth of planets between $1.7-1.9\, R_\oplus$ \citep[][Figure~\ref{fig:radval}]{Fulton_2017}. The bimodal feature is commonly known as the Radius Valley and marks the distinction between terrestrial super-Earths and larger sub-Neptunes whose bulk compositions are rich in volatiles (e.g. H-He, water, etc.). The bulk compositions of sub-Neptunes remain a mystery as measurements of these planets' masses and radii alone are insufficient to infer unique mass fractions of planetary materials due to degeneracies with interior structure models (see Section~\ref{chap1:sec2:subsec2}). Efforts to mitigate these degeneracies include the detailed characterization of the atmospheres of sub-Neptunes, for example to distinguish between low versus high molecular weight atmospheres of H-He and other, heavy volatile-rich atmospheric compositions (e.g. water), respectively \citep[e.g.][]{Benneke_2024}. Further efforts from a modeling perspective seek to understand the geochemical interactions between the atmosphere and the (possibly molten) planetary surface that may improve our understanding of these extremely common planets \citep[e.g.][]{Kite_2020,Schlichting_2022,Innes_2023}.

The origin of the Radius Valley remains a topic of debate. One class of proposed physical mechanisms that can carve out the distinction between super-Earths and sub-Neptunes is that of {\it thermally-driven atmospheric escape} whereby planets form as terrestrial cores with primordial H-He envelopes that are subsequently lost to space. Complete atmospheric escape is achieved for small, close-in planets that end up as bare rocky cores and make up the observed super-Earth population. Conversely, planets that retain the bulk of their H-He envelopes represent the sub-Neptunes in the Radius Valley. Likely heating sources that are capable of driving bulk hydrodynamic outflows include irradiation by stellar X-ray and ultraviolet photons, as in the XUV-driven photoevaporation paradigm \citep[e.g.][]{Owen_2017} and the internal cooling luminosity of the planet's core after formation, as in the core-powered mass loss paradigm \citep[e.g.][]{Gupta_2019}. These models are largely consistent with the structure of the Radius Valley around Sun-like stars, specifically with the Radius Valley's dependence on incident stellar flux, stellar mass, and age \citep{Berger_2020,Petigura_2022}. Furthermore, the location of the Radius Valley suggests that the underlying planetary cores are composed of terrestrial materials with Earth-like mass fractions\footnote{Earth-like terrestrial planets have a mass fraction of approximately two thirds in a silicate mantle plus the remaining third in an iron core.} \citep[e.g.][]{Rogers_2021}.

An alternative hypothesis for the emergence of the Radius Valley is the idea of a primordial Radius Valley that is set directly by the planet formation process. Within this class of models, super-Earths and sub-Neptunes form as separate populations whose superposition produces the observed Radius Valley. The separate formation of these planet populations may arise from distinct formation timescales with super-Earths forming in a gas-depleted environment after the dispersal of the gaseous protoplanetary disk \citep{Lopez_2018}. Alternatively, gas-poor super-Earths may form simultaneously with the gas-enveloped sub-Neptunes but with H-He accretion that is limited by the maximum isothermal envelope that can be accreted onto low-mass planetary cores \citep[$\lesssim 1-2\, M_\oplus$;][]{Lee_2022}. Yet another plausible alternative is that sub-Neptunes are not terrestrial cores enveloped in H-He gas, instead they are water worlds that formed water-rich beyond the snow line before undergoing inward migration to their present-day locations \citep{Venturini_2020}. Unlike around Sun-like stars, a planet formation origin of the Radius Valley may be favoured over sculpting by atmospheric escape around low-mass stars ($\sim 0.3-0.7\, M_\odot$) as evidenced by the shallow slope of the Radius Valley is radius-period space, where orbital period acts as a proxy for orbital distance \citep{Cloutier_2020,Ho_2024}. Models of water-rich formation can adequately explain the Radius Valley's shallow slope around low-mass stars and further predict that the Radius Valley becomes smeared out around the lowest mass red dwarfs ($\lesssim 0.3\, M_\odot$) due to the efficient migration of water world-mass planets around these stars \citep{Burn_2021,Venturini_2024}. This prediction remains to be tested as the Radius Valley around the lowest mass red dwarf stars remains to be observationally established.

\subsection{Mass-Radius and Mass-Density Relations}\label{chap1:sec2:subsec2}
Measurements of planetary radii $R_p$ using the transit technique and planetary masses $M_p$ from complementary methods 
allow astronomers to calculate the bulk densities of (assumed spherical) planets via 

\begin{equation}
\rho_p = \frac{M_p}{\frac{4\pi}{3} R_p^3}.
\end{equation}

\noindent From measurements of $M_p$, $R_p$, and $\rho_p$, the mass-radius and mass-density relations for planetary-mass objects have been computed for small terrestrial planets up to massive gas giants. The mass-radius and mass-density relations are respectively denoted $M_p-R_p$ and $M_p-\rho_p$ for concision. Empirically-calibrating these relations allows for 

\begin{itemize} 
\item detailed investigations of planetary bulk compositions,
\item the $M_p-R_p$ relation enables predictions of planetary masses for transiting planet discoveries for which masses are often unknown (pending dedicated follow-up observations).
\end{itemize}

\noindent The $M_p-R_p$ and $M_p-\rho_p$ relations have been studied extensively over the years by different groups of researchers seeking to fit the functional forms $M_p(R_p)$ (or sometimes $R_p(M_p)$) and $\rho_p(M_p)$ for planets in different mass regimes. These studies can differ in their planet sample selections, functional forms for the relations, fitting techniques, the inclusion or exclusion of planets in the solar system, and whether to model the planet population in higher dimensions (e.g. with planet temperature). A detailed comparison of different empirical relations would be a lengthy and not particularly informative endeavour, so here we focus on only the major features in the $M_p-R_p$ and $M_p-\rho_p$ relations.

\begin{figure}
\centering
\includegraphics[width=0.5\hsize]{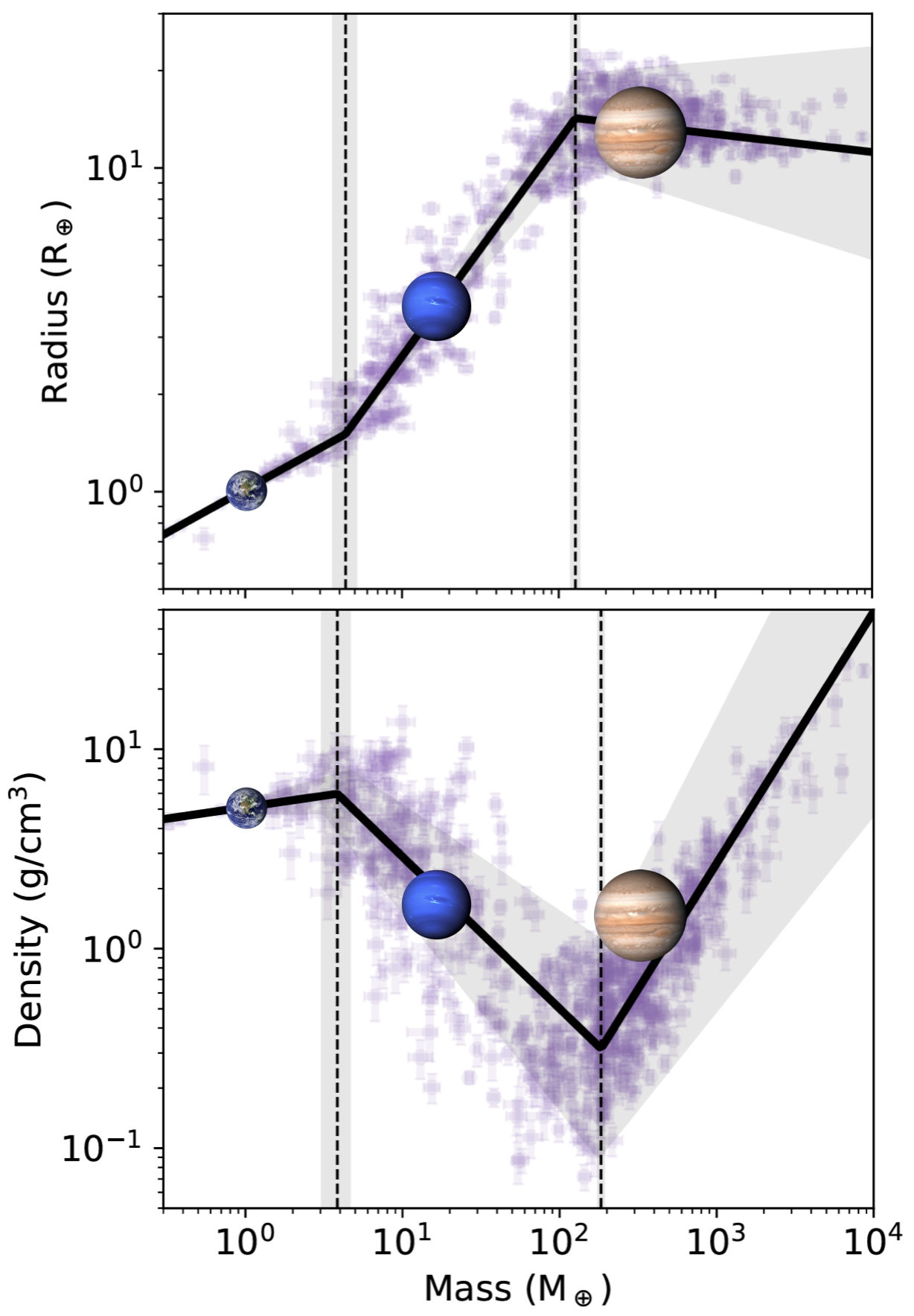}
\caption{{\bf Top:} the empirical mass-radius relation for exoplanets. The relation is fit by a broken power-law that separates three mass regimes: terrestrial worlds ($M_p \lesssim 4\, M_\oplus$), Neptunian worlds ($4 \lesssim M_p/M_\oplus \lesssim 150$), and Jovian worlds ($M_p \gtrsim 150\, M_\oplus$). The solar system planets Earth, Neptune, and Jupiter are included as illustrative cases in each planet mass regime. {\bf Bottom:} the empirical mass-density relation that helps to illustrate the physical response of different planetary materials to changes in mass (i.e. under different pressure regimes). Modified from \cite{Muller_2024}.}
\label{fig:mr}
\end{figure}

An example of the $M_p-R_p$ and $M_p-\rho_p$ relations is shown in Figure~\ref{fig:mr}. Both relations are fit by a broken power-law with the first break at the typical mass of the transition between terrestrial objects and volatile-rich Neptunes/sub-Neptunes ($\sim 4\, M_\oplus$) and the second break at the transition between Neptunian worlds and Jovian worlds ($\sim 100-200\, M_\oplus$). The terrestrial planet regime extends up to $\sim 1.8\, R_\oplus$ (i.e. the first peak in the Radius Valley; c.f. Figure~\ref{fig:radval}) and exhibits a positive scaling between planet mass and radius: $R_p \propto  M_p^{0.27}$. This scaling is shallower than if terrestrial planets followed a constant density relation (i.e. $R_p \propto M_p^{0.33}$) because at the extreme pressures found in the interiors of Earths and super-Earths, iron and silicates, the building-blocks of terrestrial planets, are compressible. The impact of compressible refractory materials is corroborated by the positive scaling in the $M_p-\rho_p$ relation for terrestrial worlds (c.f. lower panel Figure~\ref{fig:mr}). As discussed in Section~\ref{chap1:sec2:subsec1}, Neptunian worlds are volatile-rich. While the bulk of the volatile content is unknown across the Neptunian world mass regime, the $M_p-R_p$ relation of these objects is positive ($R_p\propto M_p^{0.67}$) while their $M_p-\rho_p$ relation is negative. As these planets accrete volatile-rich material and grow in mass, they also grow in size, and much more rapidly than the terrestrial planets. This is because a gaseous envelope can become hugely extended for even a small amount of gas. For example, a planet with an H-He mass fraction of just 1\% can have an envelope that doubles the size of the planet relative to its underlying solid core \citep{Owen_2017}. The rapid increase in Neptunian world radii with increasing mass also explains why their density falls off rapidly with mass. We also note that unlike the terrestrial worlds, volatile envelopes are highly temperature-sensitive, which contributes to the excess scatter in the $M_p-R_p$ and $M_p-\rho_p$ relations for Neptunian worlds in Figure~\ref{fig:mr}. As planet mass continues to increase, the compressibility of the H-He envelopes that make up the massive Neptunian worlds and the Jovian worlds begins to dictate the shape the of $M_p-R_p$ relation. A material's equation of state (EOS) describes the response of the material's density to changes in pressure and temperature (i.e. $\rho(P,T)$). At large masses $\gtrsim 180\, M_\oplus$--and correspondingly high internal pressures--the EOS of cold H-He causes massive gas giant planets to become severely compressed. This marks the transition from Neptunian to Jovian worlds wherein the density of Jovian worlds increases rapidly with increasing mass. The compressibility of H-He produces the turnover in the $M_p-R_p$ relation to a slightly negative scaling ($R_p \propto M_p^{-0.06}$) whereby Jovian planets exhibit comparable sizes over nearly two orders of magnitude in mass from about half the mass of Jupiter ($\sim 159\, M_\oplus$) into the regime of substellar objects above $\sim 4000\, M_\oplus$.

\begin{figure}
\centering
\includegraphics[width=0.8\hsize]{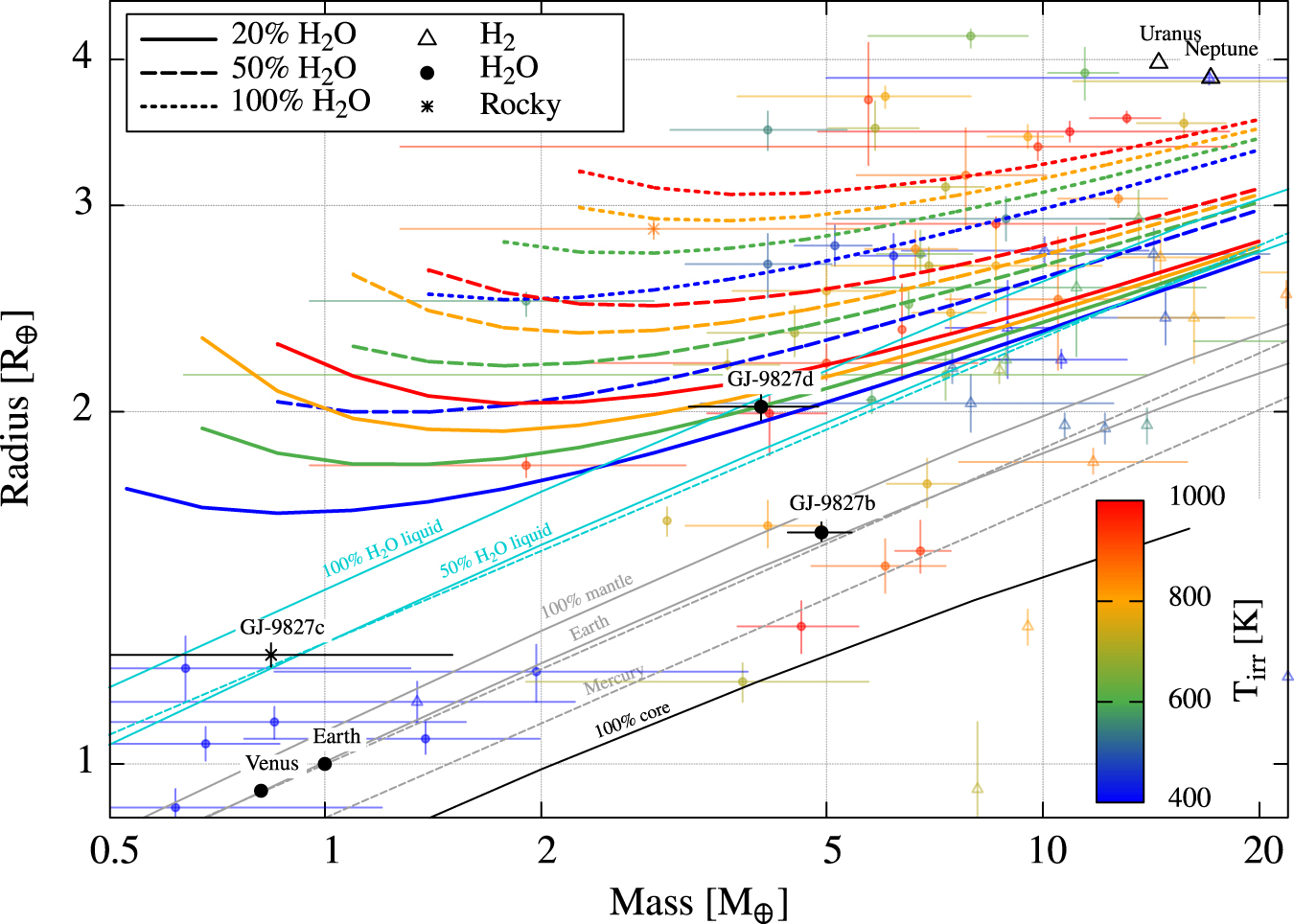}
\caption{Theoretical mass-radius relations for different planet compositions. These include solid planets with iron core mass fractions of 100\%, 70\% + 30\% MgSiO$_3$ (i.e. Mercury-like), 33\% + 67\% MgSiO$_3$ (Earth-like), 0\% (i.e. 100\% MgSiO$_3$ mantle), 50\% MgSiO$_3$ + 50\% H$_2$O, and 100\% H$_2$O. Also included are planets with water mass fractions of 20\%, 50\%, and 100\% over a range of irradiation temperatures between 400-1000 K. A select set of exoplanets with measured masses and radii and solar system planets are overplotted for comparison to the various theoretical mass-radius relations. Courtesy \cite{Aguichine_2021}.}
\label{fig:mr2}
\end{figure}

While the $M_p-R_p$ relation in Figure~\ref{fig:mr} represents the empirical relation based on exoplanets with measured masses and radii, theoretical $M_p-R_p$ relations can also be calculated for planets whose bulk compositions are a mixture of materials with known EOSs. For a hypothetical planet with a known mass and relative mass fractions of different materials with known EOSs (e.g. iron, silicates, water, H-He, etc.), the corresponding radius of the planet can be computed by numerically solving the differential equations of mass continuity and hydrostatic equilibrium. For example, a simplified view of the Earth's interior is a pure iron core with a mass fraction of $\sim 33$\%, plus a $\sim 67$\% mantle of MgSiO$_3$ perovskite with other minor refractories like Al, Ca, and Na being ignored. The EOS of iron and MgSiO$_3$ are well understood from laboratory experiments and can be used to numerically solve for the radius of a planet with an Earth-like composition of arbitrary mass. Examples of theoretical $M_p-R_p$ relations for different planet compositions, including Earth-like composition, are shown in Figure~\ref{fig:mr2}. The iron core mass fraction (CMF) of the Earth is $\approx 33$\%, while Mercury's CMF is $\approx 70$\%. The corresponding theoretical $M_p-R_p$ relations are included in Figure~\ref{fig:mr2}, along with illustrative relations for CMF $=0$\% and 100\%. Iron is a higher-density material than perovskite such that the theoretical $M_p-R_p$ relations for higher CMF values correspond to higher densities (i.e. smaller planet radii for a fixed mass). 

While the EOSs for iron and perovskite are isothermal\footnote{An isothermal EOS is one that does not have an explicit dependence on temperature.} at the typical pressures found in super-Earth interiors, the EOS of water is sensitive to temperature. This effect is demonstrated in Figure~\ref{fig:mr2} as multiple theoretical $M_p-R_p$ relations are depicted with varying water mass fractions (WMF) for irradiation temperatures\footnote{A planet's irradiation temperature $T_{\mathrm{irr}}=T_{\mathrm{eff}} \sqrt{R_\star/2a}$ is set by the radiative balance between itself and its host star located at a distance $a$ and with an effective temperature and radius of $T_{\mathrm{eff}}$ and $R_\star$, respectively. Note that $T_{\mathrm{irr}}$ is independent of the planet's Bond albedo $A_B$ (a.k.a. reflectivity), which is an unobservable quantity for the majority of known exoplanets.}  that range from 400-1000 K. Over this range of temperatures, surface water is in the vapour state and behaves much like an ideal gas with $P \propto T$. The increased pressure of hot steam atmospheres make their envelopes more vertically extended than at cooler temperatures, thus producing larger planet radii at hotter $T_{\mathrm{irr}}$ for a fixed WMF. Knowledge of the EOS of water in its many phases across the pressures and temperatures relevant to super-Earth interiors is critical to assessing the bulk water content of the hypothetical population of water worlds. Recall that water worlds are predicted by models of small planet formation and may make up the population of sub-Neptunes seen in the Radius Valley, particularly around low mass stars. By comparing the observed masses and radii of known exoplanets to theoretical $M_p-R_p$ relations helps provide some insight into the likely compositions of exoplanets as a function of planetary mass. For example, a careful inspection of Figure~\ref{fig:mr2} reveals that planets with Earth-like compositions can be as massive as $\sim 10\, M_\oplus$, which is much higher mass than the transition mass from terrestrial to Neptunian worlds suggested by Figure~\ref{fig:mr}. But it is true that planets with $M_p \in [4,10]\, M_\oplus$ span a wide range of sizes suggesting that the sub-Neptune planet population exhibits a wide diversity of WMF and H-He mass fractions. However, we caution that comparing a planet's measured mass and radius to theoretical $M_p-R_p$ relations does not uniquely determine its bulk compositions because the problem of determining the relative mass fractions of $>2$ chemical components from mass and radius measurements alone is an inherently degenerate problem. Observations of stellar abundance ratios Fe/Si and Mg/Si can help alleviate these degeneracies for terrestrial planets \citep[e.g.][]{Dorn_2017} while the distinction between water worlds and H-He-enveloped sub-Neptunes will require detailed atmospheric characterizations to infer the relative fraction of H-He gas versus heavier volatile species like water.

\subsection{Neptunian desert}\label{chap1:sec2:subsec3}
\begin{figure}
\centering
\includegraphics[width=0.95\hsize]{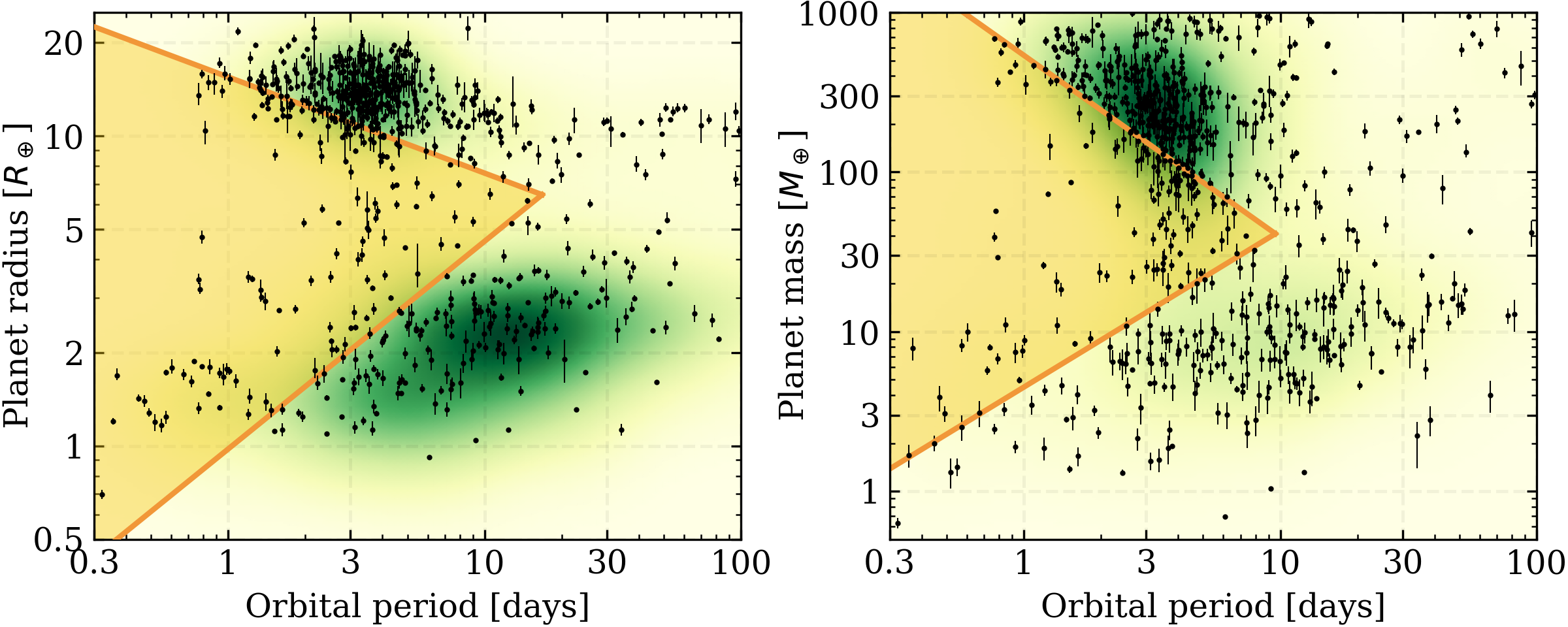}
\caption{The Neptunian desert (shaded region) in the planet radius-period and planet mass-period parameter spaces. The triangular desert boundaries are defined by \cite{Mazah_2016}. The background colourmap depicts the number density of confirmed exoplanets, which are included as black markers and were retrieved from the NASA Exoplanet Archive. Modified from \cite{Osborn_2023}.}
\label{fig:desert}
\end{figure}

The Neptunian desert, also known as the hot Neptune desert or the sub-Jovian desert, is a wedge-shaped region of the exoplanetary radius-period and mass-period spaces marked by a dearth of planets located within the desert as depicted in Figure~\ref{fig:desert}. The Neptunian desert was first discovered by \cite{Sazbo_2011} with early discoveries from the Kepler mission. Parameter limits were later defined by \cite{Mazah_2016} and roughly corresponded to planet sizes $R_p \in [2,9]\ R_\oplus$ and masses $M_p \in [10,250]\, M_\oplus$, with orbital period of $\lesssim 4$ days. Leading theories that can explain the location and triangular shape of the desert include the sculpting of the desert's lower boundary by photoevaporation, whereby XUV stellar irradiation can efficiently strip the atmospheres of close-in planets, thus reducing their masses and radii \citep{Owen_2018}. Photoevaporation presents a natural explanation for the positive slope of the desert's lower boundary in Figure~\ref{fig:desert} because the incident XUV flux is great at small orbital separations such that planetary atmospheres should become less susceptible to XUV-driven escape, and therefore appear larger, as they become progressively further away from their host stars (i.e. at longer orbital periods). Conversely, the desert's upper boundary may be the result of a tidal disruption barrier wherein giant planets are excited onto highly eccentric orbits, due to gravitational interactions with other planets or distant massive companions, before tidal dissipation circularizes their orbits close to the host star. Giant planets with masses $\gtrsim 1$ Jupiter mass, can undergo tidal decay and reach shorter orbital periods without being tidally distributed \citep{Owen_2018}. The mass dependency of the tidal disruption barrier naturally explains the negative slope of the desert's upper boundary. Planets embedded deep within the desert are stable against photoevaporative mass loss and tidal disruption.

While results from the Kepler mission suggested that the Neptunian desert was almost entirely devoid of planets, numerous recent discoveries from NASA's Transiting Exoplanet Survey Satellite (TESS) mission have begun to populate the desert. What does this mean for the emergence of the Neptunian desert? Astronomers have yet to find a definitive answer, but observational clues include the fact that these planets found wandering the desert orbit a variety of stars from low-mass red dwarfs to Sun-like stars. Many desert planets are also amenable to atmospheric characterization efforts, which may shed light on the physical processes that enable these planets to survive in the desert.

\subsection{The Peas in a Pod Pattern}\label{chap1:sec2:subsec4}

\begin{figure}
\centering
\includegraphics[width=.44\hsize]{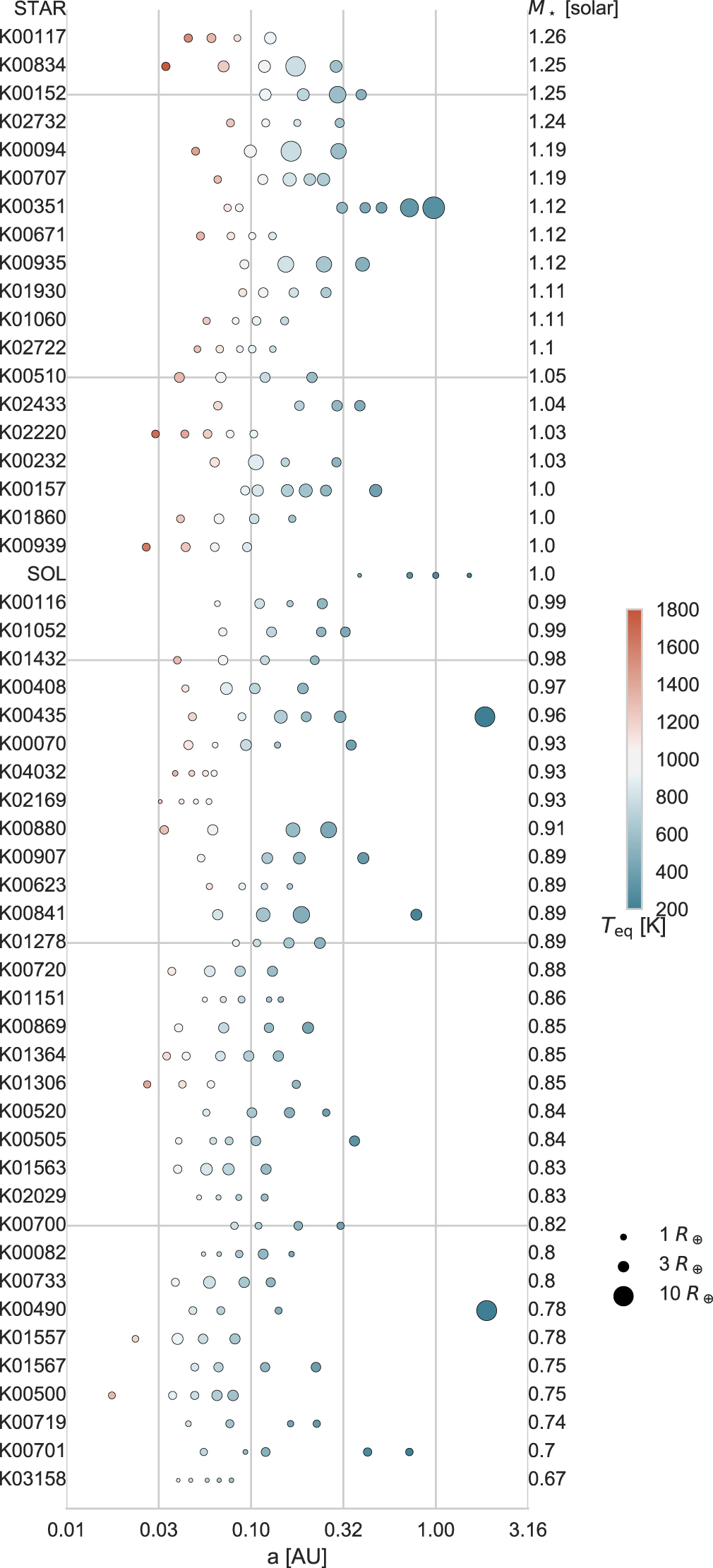}
\caption{The architectures of multi-planet systems of tightly-packed inner planets with $\geq 4$ transiting planets. Planets in these systems appear analogously to ``peas in a pod" due to their correlated sizes and orbital spacings. Courtesy \cite{Weiss_2018}.}
\label{fig:peas}
\end{figure}

Another major discovery from the Kepler space mission is that systems of tightly-packed inner planets (STIPs) appear to have correlated sizes, masses, and orbital spacings \citep{Weiss_2018}. This features is clearly visualized in Figure~\ref{fig:peas}, which depicts the architectures of Kepler STIPs with $\geq 4$ known transiting planets. Planetary systems that exhibit the peas in a pod uniformity (PiaP) present a treasure trove of features that inform our understanding of how super-Earths and sub-Neptunes form. For example, planet size is strongly correlated with the size of its neighbour. That is, every planet in a multi-planet STIP is more likely to be similar in size to its neighbouring planet than having been drawn from the full exoplanet size distribution, which spans more than an order-of-magnitude in planet radii. Furthermore, in approximately two thirds of planet pairs, the outer planet is larger than the inner planet. This configuration is consistent with expectations planet formation models of both solids and gas accretion, as well as with expectations from thermally-driven atmospheric mass loss. These intra-system trends become more significant when super-Earths and sub-Neptunes are considered separately \citep{Millholland_2021}.

The architectures of PiaP systems not only exhibit striking uniformity of planet sizes, but also in orbital spacings. High-multiplicity STIPs tend to have regular spacings between planets in terms of the orbital period ratios of adjacent planet pairs (i.e. $P_{i+1}/P_i$, where $i$ is the index a planet in a multi-planetary system). Small planets also tend to exhibit smaller period ratios than systems with larger planets. Consequently, systems of multiple known super-Earths and sub-Neptunes tend to be very compact in comparison to the solar system (c.f. the top row in Figure~\ref{fig:peas}). The uniformity of orbital spacings strongly suggests that dynamics plays a key role in setting the architecture of multi-planet systems. The PiaP pattern has been replicated by planet population synthesis models that couple planetesimal and gas accretion with evolutionary processes such as atmospheric escape and dynamical evolution \citep{Mishra_2021}. This suggests that the PiaP pattern is a natural outcome of dynamical interactions between planets following the earliest stages of the system's formation.

The PiaP pattern demonstrates the clear link between planets' physical and orbital parameters, and how these properties are linked through the planet formation process. In the next section, we will explore these and other orbital features that have been uncovered in the population of known exoplanets.

\section{Population-level trends in planets' orbital parameters}\label{chap1:sec3}
\subsection{Planet multiplicity and mutual inclinations}\label{chap1:sec3:subsec1}
Planet multiplicity refers to the number of known planets in a planetary system. For any given planetary system, this number may not be equal to the true number of planets in the system. In transiting planetary systems like those discovered by Kepler, each planet's orbital inclination relative to the observer's line-of-sight can be measured from the duration of transit. Mutual inclinations refer to the difference in orbital inclinations between planet pairs in the same planetary system, where a mutual inclination $\Delta i = 0^\circ$ implies a coplanar planet pair. The orbital inclinations of most non-transiting planets are not observationally accessible, which relegates investigations of the exoplanet mutual inclination distribution to transiting planetary systems.

High-multiplicity systems that feature at least four known planets represent fewer than 1\% of Kepler systems. Furthermore, only 22\% of Kepler systems appear to have more than one planet. Parametric planet populations and planet formation simulations that are able to reproduce Kepler's yield of singles and multis, along with the distribution of transit durations (which encodes information on planets orbital inclinations), significantly underpredicts the number of stars that host single planet systems \cite{Lissauer_2011}. The apparent excess of systems with a single transiting planets leads to the possibility of two populations of planets with either different intrinsic multiplicities or mutual inclination distributions, and is commonly referred to as the {\it Kepler dichotomy}.

While multiple solutions have been proposed to explain the Kepler dichotomy, a consensus on the true underlying cause of the dichotomy remains elusive. Proposed solutions include adopting two populations of planet multiplicities \citep{Fang_2012} or mutual inclination distributions corresponding to highly excited and coplanar distributions \citep{He_2019}. These distributions include the commonly-adopted Rayleigh distributions with scale parameters of $\sigma_{\Delta i,high} \gtrsim 20^\circ$ and $\sigma_{\Delta i,low} = 1.4^\circ$, respectively. The mutual inclinations of the latter population resemble that of the solar system ($\Delta i = 1.0^\circ$), whereas the former population of system with high mutual inclinations would represent dynamical states that are much more highly excited compared to the solar system. Compelling alternative solutions to the Kepler dichotomy that avoid dichotomous mutual inclination distributions include a single distribution that is appropriately broad and tends toward coplanarity with increasing planet multipliciity \citep{Zhu_2018,He_2020}. The apparent excess of Kepler singles may also be attributed to selection biases as opposed to an astrophysical effect because the degraded transit detection completeness in high multiplicity systems, if left unaccounted for, would contribute to exaggerating the excess of the singles in studies of Kepler's multiplicity demographics  \citep{Zink_2019}.

\subsection{Orbital spacing} \label{chap1:sec3:subsec2}
Figure~\ref{fig:demo} revealed that exoplanets span orbital periods of $\sim [0.1,10^5]$ days, which corresponds to roughly four orders-of-magnitude in orbital distances. This includes planets that orbit at a fraction of the mean Earth-Sun distance--a length scale known as an {\it astronomical unit} (au)--to planets with orbits of many tens of au. This wide range of orbital separations corresponds to planets' irradiation temperatures that range from a few thousand degrees, to temperate planets with Earth-like temperatures that may be conducive to hosting an Earth-like biosphere (see Section~\ref{chap1:sec4:subsec2}), to ultra-cold planets whose energy budgets are likely dominated by internal processes than by stellar irradiation. 

The prevalence of STIPs discovered by Kepler present very different orbital architectures than in the solar system. These systems can be understood in terms of their proximity to dynamically unstable configurations. First, we define the mutual Hill radius between a pair of adjacent planets indexed 1 and 2 as

\begin{equation}
R_{\mathrm{Hill}} = \frac{a_1 + a_2}{2} \left( \frac{M_{p,1}+M_{p,2}}{3M_\star} \right)^{1/3},
\end{equation}

\noindent where planet $i$ has a semi-major axis $a_i$, mass $M_{p,i}$, and both planets orbit a star of mass $M_\star$. A common metric for dynamical stability is the Hill stability criterion whereby the separation between the planet pair must exceed a critical number of mutual Hill radii in order to be stable over Gyr timescales:

\begin{equation}
\frac{|a_2 - a_1|}{R_{\mathrm{Hill}}} > K.
\label{eq:hill}
\end{equation}

\noindent Here, $K$ is a dimensionless number whose value is multiplicity-dependent, but is commonly set to $K=8-10$ when considering a single planet pair of similar-mass planets, as are commonly found in STIPs. 

Because transit surveys are sensitive to planet radii and not to planet masses, the Hill radii of planets studied with transit observations alone cannot be directly assessed. In practice, masses are estimated from an empirically-calibrated mass-radius relation, such as the one shown in Figure~\ref{fig:mr}. By estimating the masses of small Kepler planets in compact systems and calculating the mutual Hill radii of planet pairs in multi-planet systems, astronomers have established that the separations in high-multiplicity systems are tightly clustered around twelve mutual Hill radii, just beyond the Hill stability threshold of $K\sim 8-10$ \citep{Pu_2015}. The ubiquity of high-multiplicity systems teetering on the edge of stability suggests that compact STIPs formed with more planets and tighter spacings than what is evident in the present-day population. It is the dynamical interactions between planets during the early evolutionary stages of the system that sculpt the system's architecture by ejecting planets and increasing the orbital spacing between surviving planets.

\subsection{Orbital Eccentricities}\label{chap1:sec3:subsec3}
Another important orbital parameter describing the degree of dynamical excitation is orbital eccentricity; a measure of the ellipticity of a planet's orbit. The majority of known planets follow nearly Keplerian orbits\footnote{The exceptions being those planets that exhibit strong gravitational interactions with additional bodies in the system.}, which are elliptical by definition. An ellipse is described by the lengths of its major and minor axes, or more conventionally in orbital mechanics, the lengths of its semi-major and semi-minor axes $a$ and $b$, respectively. For orbital eccentricities corresponding to bound orbits ($a,b>0$),

\begin{equation}
e \equiv \sqrt{1 - \left( \frac{b}{a} \right)^2 }.
\end{equation}

\noindent A circular orbit is simply a special case of an ellipse for which $a=b$ such that the orbital eccentricity equals zero. Increasing $e$ is equivalent to increasing the orbit's ellipticity by shrinking the minor axis relative to the major axis. This can continue until $b=0$ at which point $e=1$ and the orbit is no longer defined; the trajectory has become an unbound parabola. Trajectories with eccentricity values that exceed unity are hyperbolic. 

\begin{figure}
\centering
\includegraphics[width=0.8\hsize]{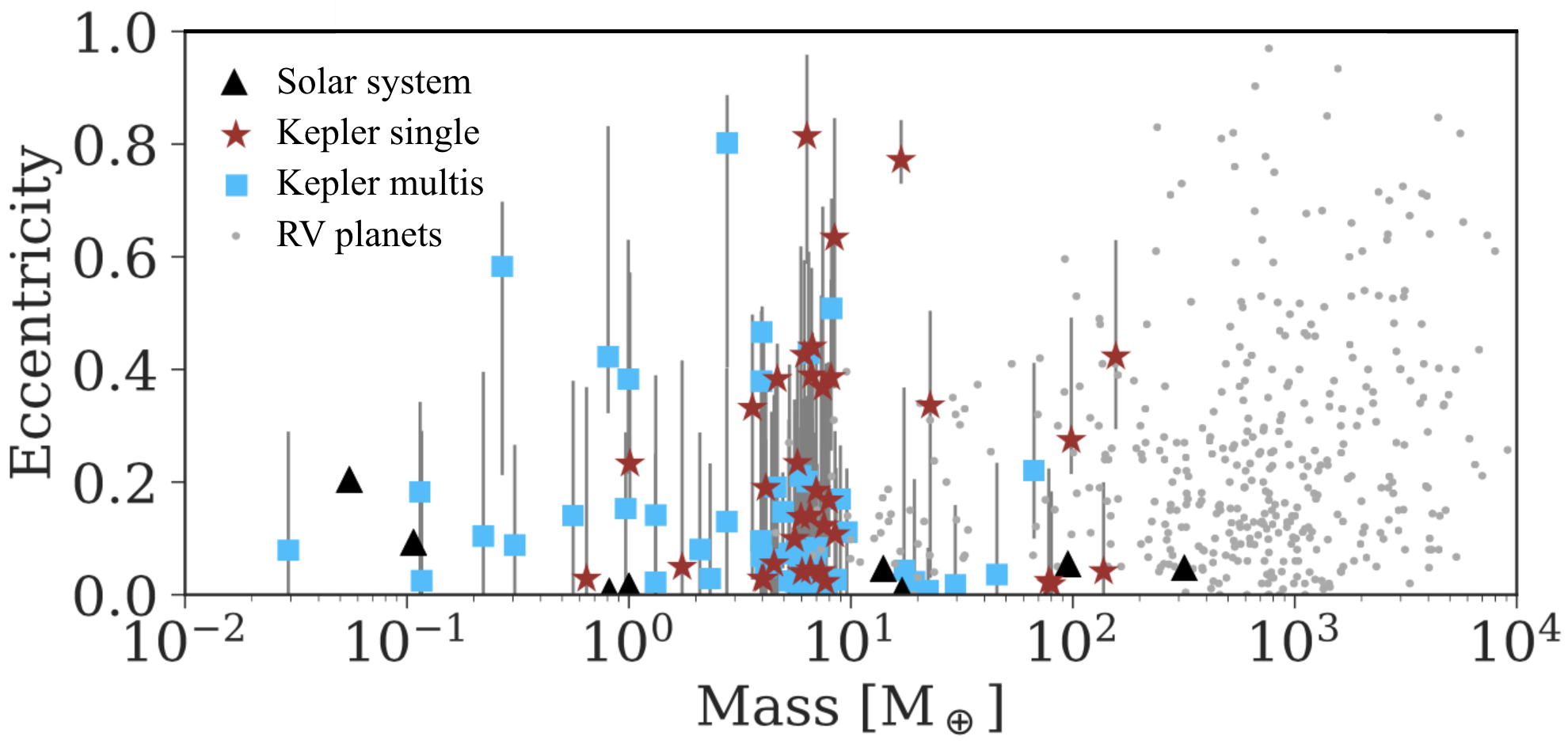}
\caption{The distribution of planet masses and orbital eccentricities for the solar system planets (triangles) and exoplanets (other markers). Eccentricity measurements from the Kepler primary mission are depicted as stars and squares for single and multi-planet systems, respectively. Results from radial velocity surveys are depicted as grey dots. Modified from \citep{VanEylen_2019}.}
\label{fig:ecc}
\end{figure}

Expectations from isolated planet formation in a protoplanetary disk are that primordial eccentricities should be nearly zero due to gas drag that dampens the excitement of orbital eccentricities. However after their formation, planets' orbital eccentricities can be excited by gravitational interactions with other massive bodies either via close encounters or by action at a distance with distant companions. The mechanism that excites the orbital eccentricities also operates on planets' orbital inclinations such that the excitement of mutual inclinations in multi-planet systems should extend to the distribution of orbital eccentricities due to energy conservation. This is empirically supported by evidence that the orbital eccentricities of small Kepler planets follow two distinct distributions depending on the system multiplicity \citep[Figure~\ref{fig:ecc};][]{VanEylen_2019}. Specifically, the eccentricity distribution among single transiting planetary systems is much broader ($\sigma_e=0.32$) than in the multi-planet systems ($\sigma_e=0.083$). The finding that planets' eccentricities in multi-transiting systems are clustered around zero is consistent with expectations from dynamical stability arguments because large orbital eccentricities in multi-planet systems increases the likelihood of an orbit crossing and, consequently, the destabilization of the system. The impact of eccentricity on dynamical stability can be evaluated by generalizing the Hill stability criterion in Eq.~\ref{eq:hill} to 

\begin{equation}
\frac{a_2(1-e_2) - a_1(1+e_1)}{R_{\mathrm{Hill}}} > K,
\label{eq:hillv2}
\end{equation}

\noindent where $e_1$ and $e_2$ are the orbital eccentricities of the inner and outer planets, respectively. The terms in the numerator of the left-hand side of Eq.~\ref{eq:hillv2} are the closest approach of planet 2 to its host star (i.e. perigee $=a_2(1-e_2)$) and the furtherest distance of planet 1 to the host star (i.e. apogee $=a_1(1+e_1)$). 

These results can be contrasted against those from radial velocity (RV) surveys, which is a powerful technique for measuring orbital eccentricities for all but the lowest signal-to-noise planets (typically low mass super-Earths and sub-Neptunes). The results from RV surveys also extend up into the mass regime of giant planets ($\gtrsim 100\, M_\oplus$) and span a wide range of orbital periods between $\sim 1$ to $\gtrsim 10,000$ days. The eccentricities of planets form a broad distribution with a significant fraction of high-eccentricity planets with $e\gtrsim 0.5$, which far exceeds the eccentricities of the planets in the solar system. The orbital eccentricity distribution serves an observational diagnostic of planets' dynamical histories. The large fraction of giant planets with excited orbital eccentricities points toward dynamical histories that commonly featured gravitational interactions with additional planets in the system and/or distant stellar or substellar companions. The resulting high-eccentricity migration (HEM) of giant planets suggests that HJ formation is consistent with ex-situ formation, plus HEM and subsequent tidal circularization that produce the prevalence of HJs on nearly circular orbits that are seen today.

\subsection{Orbital obliquities} \label{chap1:sec3:subsec4}
The inclination of the normal vector to a planet's orbital plane to the rotation axis of its host star introduces another unique diagnostic of the planet's dynamical history. This quantity is known as the orbital obliquity or the spin-orbit angle $\lambda$. Orbital obliquity measurements are possible for transiting planets through the measurement of the Rossiter-McLaughlin effect whereby a transiting planet occults different portions of the rotating star's profile that are either moving toward or away from the observer (as long as the stellar rotation axis is not pointed directly toward the observer). As the planet occults the approaching or receding limbs of the star throughout its transit, it produces a pseudo-signal in the star's apparent radial velocity that is sensitive to the projected spin-orbit angle $\lambda$. If the inclination of the stellar rotation axis is known--from photometric measurements of the stellar rotation period and projected rotation velocity $v\sin{i}$--then $\lambda$ can be deprojected to reveal the true orbital obliquity $\psi$. Measurements of $\lambda$, or ideally $\psi$, allow for the identification of aligned prograde orbits ($\lambda=0^\circ$), misaligned prograde orbits ($\lambda \lesssim 45^\circ$), polar orbits ($\lambda=90^\circ$), and retrograde orbits ($\lambda > 90^\circ$), thus tracing the prevalence of HEM mechanisms, which can also excite orbital obliquities. 

\begin{figure}
\centering
\includegraphics[width=\hsize]{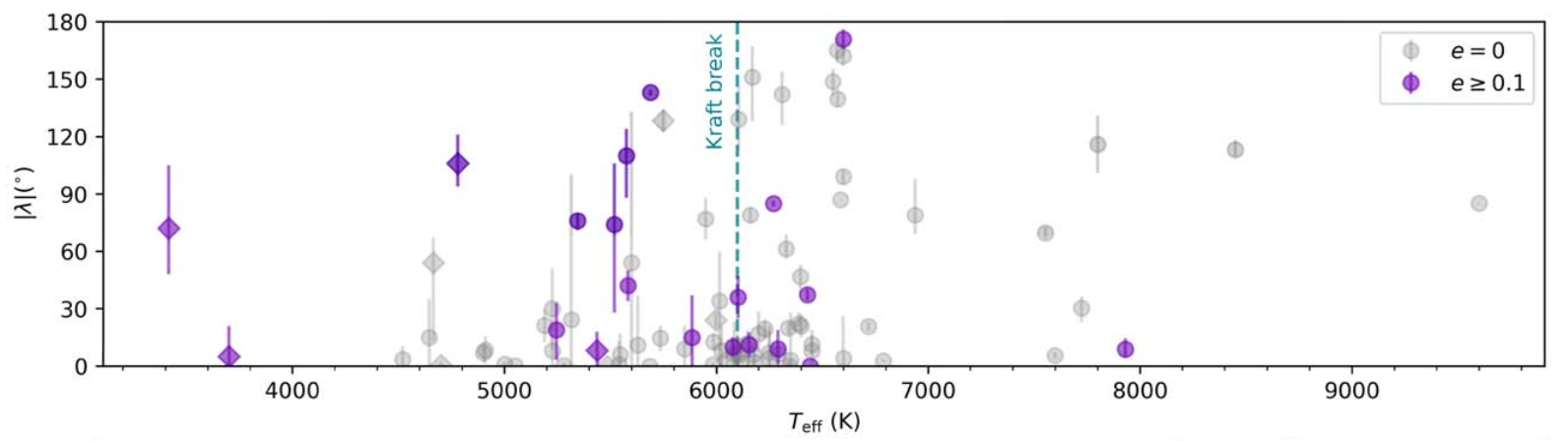}
\caption{Measurements of the projected orbital obliquity angle $\lambda$ versus host stellar mass for eccentric ($e\geq 0.1$) and non-eccentric planets ($e=0$). Most of the planets herein with orbital obliquity measurements are HJs, although a handful of sub-Saturn-mass planets are also included. The population of close-in exoplanets exhibit a diverse set of obliquity angles from aligned ($\lambda=0^\circ$), to misaligned ($\lambda \lesssim 45^\circ$), polar ($\lambda=90^\circ$), and retrograde orbits ($\lambda > 90^\circ$). Planets orbiting hot stars above the Kraft break experience weaker tidal damping and thus, elevated orbital obliquities relative to comparable planets that orbit cooler stars. Courtesy \citep{Rice_2022}.}
\label{fig:obl}
\end{figure}

The distribution of planets' projected orbital obliquities as a function of host stellar mass is shown in Figure~\ref{fig:obl}. Most of the planets shown therein are gas giants, although there are a few examples of hot Neptunes with Rossiter-McLaughlin measurements. Eccentric ($e \geq 0.1$) and non-eccentric ($e=0$) planets are considered separately to assess the role that eccentricity plays in planets exhibiting a wider range of obliquities around stars hotter than the Kraft break at $\sim 6100$ K, compared to around cooler stars. Stars cooler than the Kraft break have thick outer convective layers that help drive their large-scale magnetic dynamos, which in-turn drive their angular momentum evolution through the effect of magnetic breaking\footnote{Magnetic breaking refers to magnetized stellar winds that are able to carry away the stars rotational angular momentum, thus slowing the stellar spin rate over time.}. Stars hotter than the Kraft break lack the outer convective layers needed to drive magnetic breaking and consequently rotate more rapidly than the cooler Sun-like stars. Stellar rotation effects the timescale for tidal realignment of orbital obliquities such that the distribution of $\lambda$ values is expected to vary across the Kraft break \citep{Albrecht_2012}. Planets that orbit these hot stars have been shown to exhibit a wider range of orbital obliquities $\lambda$, although this is only exhibited by planets on circular orbits  \citep{Rice_2022}. This result is expected as a natural outcome of HEM plus tidal damping of excited obliquities as circular planets around stars below the Kraft break experience the strongest obliquity damping. The consistency of giant planet orbital obliquities in Figure~\ref{fig:obl} with theoretical expectations from HEM plus tidal damping is corroborated by an apparent excess of warm/cold Jupiters in HJ systems, suggesting that HEM via gravitational interactions with distant planetary-mass companions (i.e. Lidov-Kozai migration) may be the dominant driver of HJ formation \citep{Zink_2023}, without the need to invoke disk-driven migration or in-situ formation to explain the properties of HJs.

\section{Population-level trends with host stellar parameters}\label{chap1:sec4}
\subsection{Exoplanet demographics over time}\label{chap1:sec4:subsec1}
The vast majority of known exoplanets in Figure~\ref{fig:demo} exist around what are known as {\it field stars}. Due to dynamical mixing, the galactic kinematics of field stars make them untraceable back to the stellar clusters in which they formed. Without the luxury of using well-established cluster ages used to date the ages of field stars, alternative techniques applicable to field stars need to be applied. One such technique that is commonly employed in the field is {\it gyrochronology}, which uses measurements of stellar rotation periods to compare to spin-down laws and infer field star ages, which typically range from one to a few Gyrs. Our understanding of the exoplanet population is therefore largely a snapshot taken during the intermediate stages of a planetary system's lifetime, after the system's early evolution but before stellar evolutionary processes alter the planetary system (e.g. planetary engulfment). 

The physical and orbital properties of individual planets and the planetary systems in which they reside are expected to undergo substantial evolution over the course of their lifetimes. This is especially true during the first few hundred Myrs following the system's formation. Examples of evolutionary drivers include planetary migration whereby gravitational interactions between planets with either their natal disks, distant stellar companions (i.e. Lidov-Kozai migration), or other planets within the system, cause planets to undergo radial migration that may be inward or outward. Multiple lines of empirical evidence point toward the prevalence of planet migration.This includes the existence of resonant chains in select multi-planet systems that must have formed via convergent disk-driven migration whereby the presence of the gas dampens forced eccentricities and helps to maintain the system's stability. Other lines of evidence for planet migration include the abundance of dynamically hot close-in planets with elevated eccentricities and orbital obliquities that are believed to result from a HEM mechanism, as well as retrievals of C/O ratios and refractory:volatile ratios (e.g. K/O) in HJ atmospheres \citep{Feinstein_2023}, which serve as signatures of planet formation in the outer regions of protoplanetary disks \citep[e.g.][]{Oberg_2011}.

One of the physical processes that alter planetary radii include radius inflation of hot gas giants that results from additional heating sources beyond stellar irradiation. This may include tidal heating after HEM, thermal tides, or the deposition of energy from irradiation into the planet's interior. The radii of gas giants and sub-Neptunes alike can also decrease with time due to any combination of thermal contraction of the planet's atmosphere after it decouples from the protoplanetary disk gas, as well as atmospheric mass loss that may be driven by thermal (e.g. XUV photoevaporation) or non-thermal processes (e.g. sputtering). The effects of atmospheric escape and thermal contraction should be imprinted on the exoplanet population as a function of age with smaller planets representing a larger fraction of the overall planet population at field ages compared to within the first few hundred Myrs of planetary systems' lifetimes. 

\begin{figure}
\centering
\includegraphics[width=0.8\hsize]{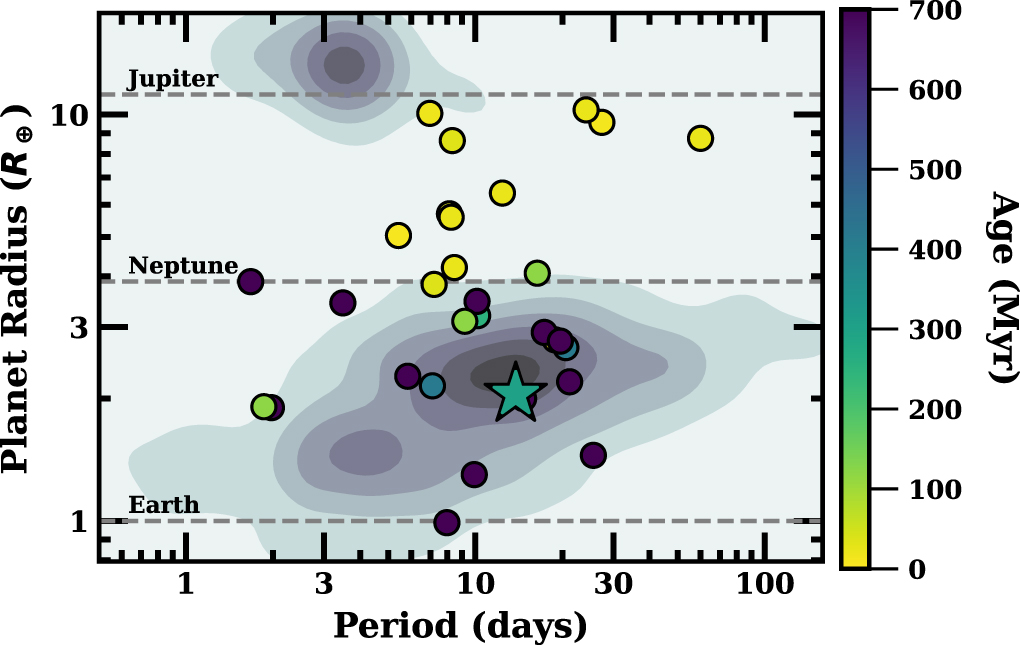}
\caption{The radii, orbital periods, and ages of a selection of confirmed young exoplanets ($<1$ Gyr). The background contours depict the full population of known exoplanets recovered by the Kepler and K2 missions, which are predominantly older than the young planets depicted by the circle and star markers (i.e. $>1$ Gyr). Courtesy \cite{Newton_2022}.}
\label{fig:thyme}
\end{figure}

The population of confirmed transiting planets around stars with well-determined ages from open star cluster membership is shown in Figure~\ref{fig:thyme}. A couple of features emerge that are suggestive of planetary formation/evolutionary processes. The absence of young HJs in the upper left portion of Figure~\ref{fig:thyme} suggests that the timescale for inward migration is longer than disk-driven migration timescale, which is constrained to the lifetime of the gaseous protoplanetary disk (i.e. within the first 10-20 Myrs). This may be interpreted as supporting evidence for an alternative dominant mechanism of HJ migration such as planet-planet scattering or secular Lidov-Kozai interactions, both of which can continue to operate much later into the system's lifetime. It is also apparent that the youngest confirmed planets $\lesssim 100$ Myrs are systematically larger than their older counterparts ($0.1-1$ Gyrs), even if the latter planets are still considered `young' in comparison to the bulk of the exoplanet population at field ages. This trend may be the result of atmospheric escape that operates on a timescale of $\sim 100$ Myrs--as is the case for XUV-driven photoevaporation around Sun-like stars--but we caution that this trend may be the direct result of observational bias because young stars exhibit high levels of magnetic activity that produce photometric variability, thus making it more difficult to detect transiting planets in photometry around young stars. That is, the planets younger than $\sim 100$ Myrs in Figure~\ref{fig:thyme} likely do not represent the full planet population at those ages, only the largest planets that are most easily detected. While there is preliminary evidence that sub-Neptunes may outnumber terrestrial super-Earths at young ages \citep{Berger_2020,Vach_2024}, more robust observational diagnostics will require more planet detections around young stars and/or measurements of atmospheric fractionation to assess the importance of atmospheric escape processes \citep{Cherubim_2024}.

\subsection{Habitable zone}\label{chap1:sec4:subsec2}
The primary science goal of the Kepler space mission was to find Earth-sized planets in the habitable zones (HZ) of Sun-like stars. While many definitions of the circumstellar HZ exist throughout the literature, their commonality is that the HZ refers to the range of orbital distances for which a terrestrial-mass planet with an N$_2$--H$_2$O--CO$_2$-dominated atmosphere can sustain liquid water on its surface. Alternative definitions include additional physical effects such as cloud feedback, orbital eccentricity, and the impact of tidal locking, all of which are expected to influence the potential for a planet to host an Earth-like biosphere, but do not differ in their fundamental requirement for sustained liquid water on the planet's surface. While the validity of the simplistic HZ model from \cite{Kopparapu_2013} is heavily debated, it is worth noting that the aforementioned effects, as well as many other planetary and stellar attributes\footnote{E.g. stellar XUV activity history, planetary magnetic field strength, tectonic activity, axial tilt to drive seasonality, a natural satellite to drive tides, etc.} will affect habitability, but many of these attributes are observationally inaccessible. This necessitates the adoption of an admittedly simplistic prescription of the circumstellar HZ based solely on the observable parameters of exoplanet size, mass, orbital separation, and eventually, atmospheric chemical composition, which remains to be successfully demonstrated on an Earth-sized exoplanet.
 
Stars throughout our galaxy exhibit a wide range of sizes and temperatures from the lowest mass red dwarf stars, that are just capable of undergoing nuclear fusion in their cores\footnote{The ability to ignite hydrogen fusion in the core is the criterion that distinguishes bona fide stars from substellar objects.}, to the extremely rare O-type stars whose intense UV luminosities and stellar winds act as critical feedback processes during star formation. The broader distributions in these stellar parameters result in stellar luminosities that span ten orders-of-magnitude. To first-order, it is the stellar luminosity that sets the orbital distances at which an orbiting planet's irradiation temperature may be conducive to hosting liquid surface water. Therefore, the location of the circumstellar HZ is a sensitive function of stellar luminosity. We note however that while a planet's irradiation temperature is set solely by the luminosity of its host star and its orbital distance, a planet's irradiation temperature differs from its surface temperature in the presence of an atmosphere, and it is the surface temperature that astrobiologists are interested in when attempting to evaluate the suitability of an exoplanet to hosting liquid water on its surface. Calculating a planet's surface temperature for a given irradiation temperature is dependent on the atmospheric properties and the incident stellar spectral energy distribution. 

A model that is commonly used throughout the literature to calculate the location of the circumstellar HZ is that of \cite{Kopparapu_2013}. This model is based on a one dimensional radiative-convective, cloud-free climate model around stars with effective temperatures ranging from 2600-7200 K, which correspond to the lowest mass red dwarfs ($0.08\, M_\odot$) to massive Sun-like stars up to $\sim 1.6\, M_\odot$.  For reference, the solar effective temperature is 5780 K. The parametric model of the effective stellar fluxes that correspond to HZ edges has the functional form

\begin{equation}
S_\mathrm{eff} = S_{\mathrm{eff},\odot} + \sum_{i=1}^{4} c_i\, T_\star^i,
\label{eq:hz}
\end{equation}

\noindent where $T_\star = T_\mathrm{eff} - 5780$ K, $T_\mathrm{eff}$ is the stellar effective temperature, and the coefficients $\{ S_{\mathrm{eff},\odot}, c_1, c_2, c_3, c_4 \}$ are tabulated by \cite{Kopparapu_2013} for four definitions of the HZ edges. These edges, which correspond to conservative versus optimistic definitions of the HZ (i.e. narrow versus wide), are detailed below.

\begin{itemize}
\item {\bf Recent Venus} ({\it optimistic inner edge} located at 0.75 au around the Sun): an empirically-motivated definition of the inner edge of the HZ based on the assumption that Venus hosted liquid water on its surface prior to $\sim 1$ Gyr ago.  
\item {\bf Moist greenhouse} (a.k.a. water-loss limit; {\it conservative inner edge} located at 0.99 au around the Sun): near the inner edge of the HZ, the relatively high temperatures imply an atmosphere rich in water vapour. Upon shrinking the orbital distance further, thus increasing the instellation flux received by the planet, water vapour is photolyzed by stellar UV photons that break up H$_2$O molecules into their constituent hydrogen and oxygen components. The hydrogen is preferentially lost to space, thus preventing the recombination of H$_2$O molecules following some cooling process. The loss of water alone is catastrophic to a planet's potential for habitability, but water loss is compounded by the prevention of condensible water cloud formation, which removes the cooling feedback of clouds and further enhances heating. The Moist greenhouse (GH) is commonly used to define the inner edge of the circumstellar HZ of exoplanetary systems.  
\item {\bf Maximum greenhouse} ({\it conservative outer edge} located at 1.70 au around the Sun): near the outer edge of the HZ, the relatively low temperatures force water to condense out of the atmosphere, leaving behind a CO$_2$-dominated atmosphere akin to present-day Mars. CO$_2$ acts as a strong GH gas by absorbing longwave radiation emitted by the planet's surface before reemitting it and insulating the planet. The GH effect helps maintain a temperate surface temperature even as the planet moves further outward to lower instellation flux levels. However with enough CO$_2$, the atmosphere becomes opaque to its own outgoing longwave radiation such that the introduction of additional CO$_2$ from processes such as volcanism, no longer contribute to further heating. That is, there is a maximum CO$_2$ GH effect. Upon increasing the planet's orbital distance beyond the HZ's outer edge, further cooling causes the CO$_2$ to condense out of the atmosphere, causing the GH effect to effectively shut off and the planet's surface temperature to catastrophically cool. The Maximum GH limit is commonly used to define the outer edge of the circumstellar HZ of exoplanetary systems. 
\item {\bf Early Mars} ({\it optimistic outer edge} located at 1.77 au around the Sun): another empirically-motivated definition of the outer edge of the HZ based on the overwhelming evidence that Mars hosted liquid surface water as early as 3.8 Gyrs ago.  
\end{itemize}

\begin{figure}
\centering
\includegraphics[width=0.95\hsize]{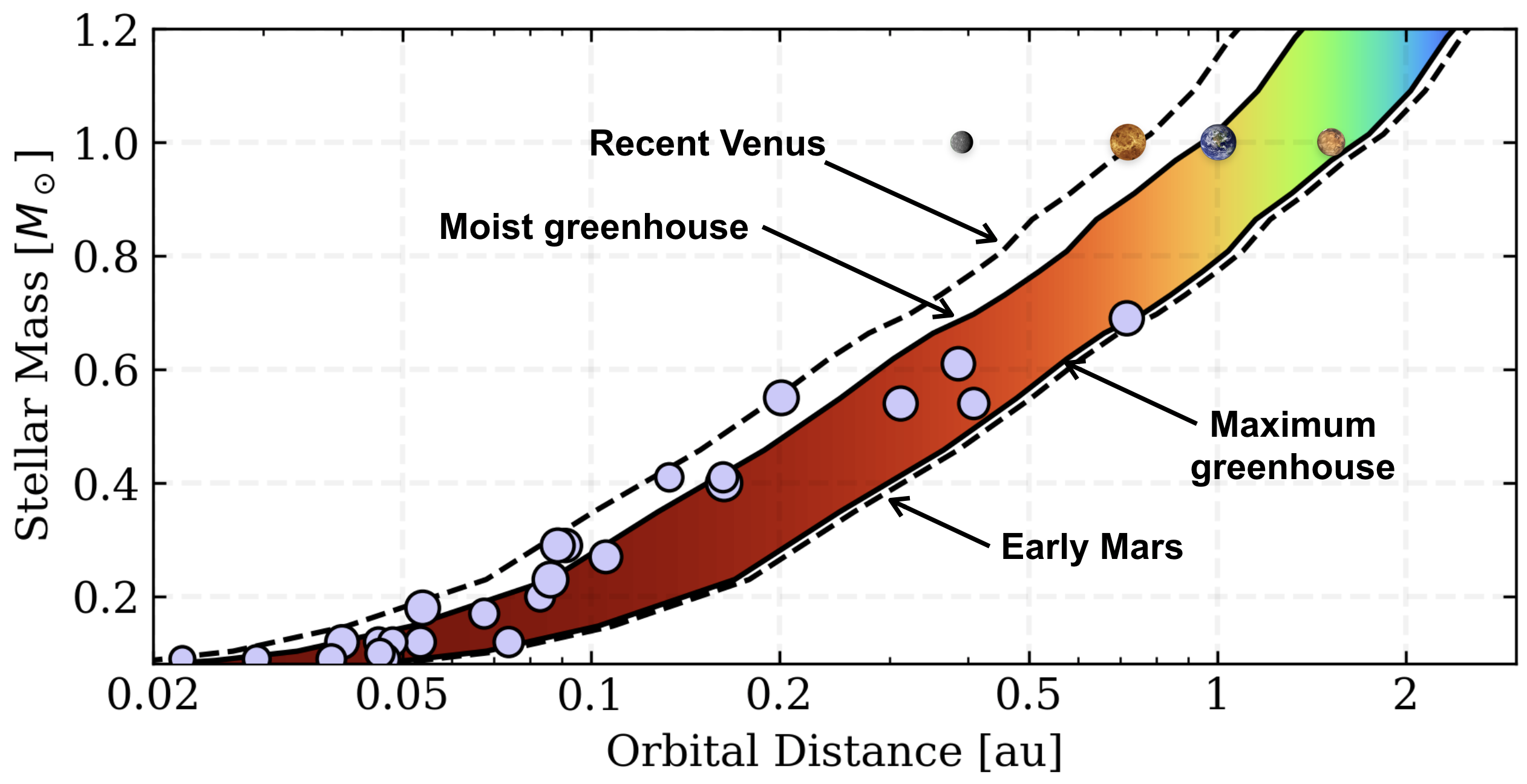}
\caption{The orbital distance limits of the circumstellar habitable zone (HZ) as a function of stellar mass based on the models from \cite{Kopparapu_2013}. The dashed lines represent the inner and outer edges of the optimistic (i.e. wide) HZ based on the Recent Venus and Early Mars definitions (see text). Similarly, the solid lines depict the inner and outer edges of the conservative (i.e. narrow) HZ based on the Moist GH and Maximum GH definitions (see text). The planets in the inner solar system are included along with confirmed exoplanets with (probable) terrestrial compositions that orbit their host stars near the HZ.}
\label{fig:hz}
\end{figure}

Given one of the definitions of a HZ edge above, the orbital distance to that HZ edge $d$ can be calculated for a star with effective temperature $T_\mathrm{eff}$ and luminosity $L_\star$ using Eq.~\ref{eq:hz} and the expression $d = 1\, \mathrm{ au}\, \sqrt{L_\star / (L_\odot\, S_{\mathrm{eff}})}$. For all the HZ definitions outlined above, the orbital distance to either edge of the HZ increases with increasing stellar luminosity, and therefore with increasing stellar mass. The effect is illustrated in Figure~\ref{fig:hz}, which shows the dependence of the distance to the moist and maximum GH edges on stellar mass. It is clear that circumstellar HZs exist much closer to low mass stars than they do around more massive stars like our Sun. This fact, combined with the bulk of known exoplanets orbiting their stars with short orbital periods (c.f. Figure~\ref{fig:demo}), means that the majority of confirmed HZ exoplanets orbit low-mass red dwarfs. Turning to the exoplanet population, there are currently $\approx 30$ confirmed exoplanets with sizes or minimum masses in the super-Earth regime that orbit within the HZ (see Figure~\ref{fig:hz}). The bulk these planets orbit low-mass red dwarfs with masses $\lesssim 0.6\, M_\odot$. A famous example is the TRAPPIST-1 system of seven terrestrial planets orbiting an ultra-cool dwarf star with $M_\star = 0.8\, M_\odot$, with 3-4 planets orbiting within the star's HZ (depending on the adopted HZ definition). Despite the design of the Kepler space telescope and its mission to measure the frequency of Earth-analogs through the galaxy, astronomers have yet to discover a true Earth analog orbiting a Sun-like star at close to 1 au. Fortunately, this may change in near future with the forthcoming launch and operation of the Nancy Grace Roman Space Telescope: NASA's next major exoplanet hunter that is slated to discover Earth-analogs using the exoplanet detection technique of gravitational microlensing.

\section{Conclusions}\label{chap1:sec5}
The more than \nplanets{} exoplanet discoveries have given rise to the field of Exoplanet Demographics: the population-level study of planetary and host stellar parameters that is unveiling the dominant formation and evolutionary processes that sculpt planetary systems throughout our galaxy. Demographics studies have been largely driven by results from NASA's Kepler space mission, which continues to produce new results more than five years after its decommissioning. The field is still considered to be in its infancy as a multitude of new facilities are slated to uncover thousands of new exoplanets that will push astronomers closer to mapping out the global architectures of planetary systems, including ones akin to our own solar system. Examples of future facilities include forthcoming data from the European Space Agency's (ESA) Gaia mission, which has been operating since 2013 and whose upcoming data release will open a window into long-period planets around stars throughout the solar neighbourhood. Looking forward, NASA's Nancy Grace Roman and ESA's PLATO missions will continue to expand upon Kepler's discovery space by enabling deep demographic studies of Kepler-like and solar system-like systems within the next few years\footnote{The anticipated launch dates for the PLATO and Roman space telescopes are 2026 and 2027, respectively.}. These discovery missions will ultimately be complemented by ESA's Ariel mission\footnote{Anticipated launch date in 2029.}. Ariel will be the first mission to conduct demographics-level studies of exoplanetary atmospheres to reveal the atmospheric properties and formation histories of exoplanets from gas giants, to the most common planets in our galaxy, the curious sub-Neptunes.


\bibliographystyle{Harvard}
\bibliography{reference}

\end{document}